\documentclass[a4paper,12pt]{article}
\usepackage[utf8]{inputenc}

\usepackage[
    backend=biber,
    style=ieee,
  ]{biblatex}
\usepackage{braket}
\usepackage{mathtools} % collection de symboles mathématiques
\usepackage{amssymb} % collection de symboles mathématiques
\usepackage{amsmath}
\usepackage{bbold}
\usepackage{amsfonts} %symboles math 
\usepackage{stmaryrd} 
\usepackage{hyperref}
\usepackage{color}
\usepackage{geometry}
\usepackage{authblk}
\usepackage[font={small}]{caption}

\newcommand{\B}{\mathcal{B}}

\newcommand{\Aut}{\mathrm{Aut}}

\newcommand{\stirlingi}{\genfrac{[}{]}{0pt}{}}

\begin{document}

\title{A Solvable Tensor Field Theory}
\author{R. Pascalie\thanks{romain.pascalie@u-bordeaux.fr}}

\affil{\small Universit\'e de Bordeaux, LaBRI, CNRS UMR 5800,  Talence, France, EU\\
\small Mathematisches Institut der Westf\"alischen Wilhelms-Universit\"at,  M\"unster, Germany, EU}

\maketitle

\begin{abstract}
We solve the closed Schwinger-Dyson equation for the 2-point function of a tensor field theory with a quartic melonic interaction, in terms of Lambert's W-function, using a perturbative expansion and Lagrange-B\"{u}rmann resummation. Higher-point functions are then obtained recursively.
%\keywords{Schwinger-Dyson equation \and series expansion and resummation \and tensor field theory}
\end{abstract}

\section{Introduction}
\label{sec:1}

Tensor models have regained a considerable interest since the discovery of their large $N$ limit
(see \cite{Gurau:2009tw}, \cite{Carrozza:2015adg}, \cite{Benedetti:2017qxl} or the book \cite{book}). 
Recently, tensor models have been related in \cite{Witten} and \cite{Klebanov:2016xxf}, to the Sachdev-Ye-Kitaev model \cite{sy}, \cite{kitaev}, \cite{Maldacena:2016hyu}, \cite{Gross:2017aos},  which is a promising toy-model for understanding black holes through holography (see also \cite{Gurau:2016lzk}, \cite{Bonzom:2017pqs}, the lectures \cite{Klebanov:2018fzb} and the review \cite{Delporte:2018iyf}).

In this paper we study a specific type of tensor field theory (TFT) \footnote{Not to be confused with tensor fields living on a space-time such as in \cite{Benedetti:2018goh}}.
More precisely, we consider a $U(N)$-invariant tensor models whose kinetic part is modified to include a Laplacian-like operator (this operator is a discrete Laplacian in the Fourier transformed space of the tensor index space). This type of tensor model has originally been used to implement renormalization techniques for tensor models (see \cite{BenGeloun:2011rc}, the review \cite{Carrozza:2016vsq} or the thesis \cite{Carrozza:2013mna} and references within) and has also been studied as an SYK-like TFT \cite{BGR}. Recently, the  functional  Renormalization Group  (FRG) as been used in \cite{Eichhorn:2018ylk} to investigate the existence of a universal continuum limit in tensor models, see also the review \cite{Eichhorn:2018phj}. This is also closely related to the Polchinski's equation for TFT \cite{Krajewski:2016svb}. Our study provides a complementary non-perturbative tool to these two approaches. 

\medskip

The Ward-Takahashi identity (WTI) for TFT, first appeared in \cite{DineWard} and has been fully established in \cite{Perez-Sanchez:2016zbh}. It was used in \cite{Sanchez:2017gxt} to derive the tower of exact Schwinger-Dyson equations (SDE) with connected boundary graph. Their large $N$ limit was established in \cite{Sanchez:2017gxt}. Then the tower of SDE with a disconnected boundary graph was derived in \cite{Perez-Sanchez:2018qkd}. Numerical methods were used in \cite{Samary:2014oya} for a $\phi^4_5$ just renormalizable tensor model to study the solutions of closed SDE for the $2$- and $4$-point functions.

Let us also mention here that the  WTI has been already successfully used to study the SDE in the context of matrix models of non-commutative quantum field theory - see \cite{Disertori:2006nq} and \cite{Grosse:2012uv}. In particular, the closed SDE 2-point function for the non-commutative and 2 dimensional $\lambda\phi^4$ has been solved in \cite{Panzer:2018tvy} using and resumming a perturbative expansion. The building block of this solution is the Lambert-$W$ function. 

\medskip

Our paper is organised as follows. In the following section we describe the setup of our work, namely the action of the model, the boundary graph expansion of the free energy and the 2-point function SDE in the large $N$ limit.
The third section is dedicated to the analysis of the perturbative expansion of the 2-point function which leads us to consider the model with one quartic melonic interaction. In the fourth section we perform the resummation of the perturbative expansion, in order to obtain the non-perturbative solution of the SDE. We then discuss shortly the higher-point functions before giving some concluding remarks. In the appendix we obtain recurrence relations on the number appearing in the perturbative expansion which translate into formulas involving Stirling numbers.

\section{The model}
\label{sec:2}

Let us consider a complex rank-$3$ bosonic tensor field theory with an action of the form
\begin{equation}\label{action}
    \mathcal{S}[\varphi,\bar{\varphi}] = \sum \limits_{\mathbf{x}}
\bar{\varphi}^{\mathbf{x}}(1+|\mathbf{x}|^2)\varphi^{\mathbf{x}} + \frac{\lambda}{N^2} \sum \limits_{c=1}^3 \sum \limits_{\mathbf{a},\mathbf{b}}\bar{\varphi}^{\mathbf{a}}\varphi^{\mathbf{b}_{\hat{c}}a_c}\bar{\varphi}^{\mathbf{a}_{\hat{c}}b_c}\varphi^{\mathbf{a}},
\end{equation}
with $\mathbf{x}=(x_1,x_2,x_3) \in \{ \frac{1}{N},\frac{2}{N}, \hdots, 1 \}^3$, $|\mathbf{x}|^2 = x_1^2 + x_2^2 +x_3^2$ and $\mathbf{a}_{\hat{c}}b_c =  (a_1, \hdots,a_{c-1},b_c,a_{c+1}, \hdots,a_D)$ for a $D$-tuple. Here $\varphi$ is a rank-$3$ bosonic tensor. Note that the quartic melonic interaction terms in the action, also called pillows, are invariant under $\mathrm U(N)^3$. The tensor fields transform as
\begin{equation}
    \varphi^{\mathbf{x}} \rightarrow \varphi^{\mathbf{x}} = \sum\limits_{y_c}U^{(c)}_{x_cy_c} \varphi^{\mathbf{x}_{\hat{c}}y_c}, \quad
    \bar{\varphi}^{\mathbf{x}} \rightarrow \bar{\varphi}^{\mathbf{x}} = \sum\limits_{y_c}\bar{U}^{(c)}_{x_cy_c}\bar{\varphi}^{\mathbf{x}_{\hat{c}}y_c},
\end{equation}
for $U^{(c)} \in \mathrm{U}(N)$ and for each colour $c \in \{ 1,2,3\}$. Each copy of the group $\mathrm{U}(N)$ acts on only one index of the tensor. Thus, the indices of the tensors have no symmetries and only indices of the same colour can be contracted.

Let us emphasise that the kinetic term above represents the discrete Laplacian in the Fourier transformed of the tensor index space and a mass term which regularises the IR divergences.

The generating functional of the model is
\begin{equation}
     \mathrm{Z}[J,\bar{J}] = \int \mathcal{D}\varphi\mathcal{D}\bar{\varphi}
     \exp{\left(-\mathcal{S}[\varphi,\bar{\varphi}] + \sum_{\mathbf{x}}
       (\bar{J}_{\mathbf{x}}\varphi^{\mathbf{x}} +
       J_{\mathbf{x}}\bar{\varphi}^{\mathbf{x}}) \right)}.
\end{equation}

\begin{figure} 
\centering
\includegraphics[scale=1]{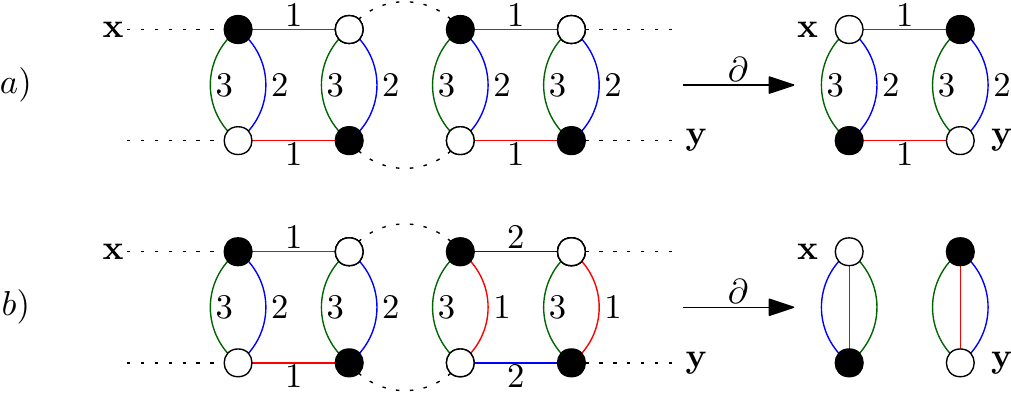}
\caption{Two connected Feynman graphs and the associated boundary graphs in the tensor field theory \eqref{action}. White and black vertex in a boundary graph $\mathcal{B}$, correspond in $\mathbb J(\mathcal{B})$ to the sources $J$ and $\bar{J}$ respectively. In the figure a) the boundary graph $V_1$ is connected and in fig. b) the boundary graph $\mathrm{m}|\mathrm{m}$ is disconnected.
%Two connected Feynman graphs and the associated boundary graphs. Dotted lines correspond to the free propagator and solid lines to the contraction of the indices of the tensors. To each external leg of a Feynman diagram is associated an external vertex so that the open graph is bipartite. These vertices are exactly the vertices of the boundary graph. An edge of colour $c$ in the boundary graph, corresponds to a path between two external leg in the Feynman graph, which alternates between dotted lines and lines of colour $c$. White and black vertex in a boundary graph $\mathcal{B}$, correspond in $\mathbb J(\mathcal{B})$ to the sources $J$ and $\bar{J}$ respectively. In the figure a) the boundary graph $V_1$ is connected. In fig. b) the boundary graph $\mathrm{m}|\mathrm{m}$ is disconnected.
\label{fig:boundary}}
\end{figure}

In tensor models, Feynman graphs (see Fig. \ref{fig:boundary}) can be drawn with two types of lines: dotted lines representing the propagator and solid lines which correspond to the contractions of the index of the tensors in the interaction. Hence, each solid line has a colour which correspond to the contracted index of the tensor. A colouring of a graph is then an edge-colouring where the solid lines have colours in $\{1,2,3\}$ and the dotted lines have the colour $0$. The Feynman graphs are then $4$-coloured graphs. We consider a complex tensor field theory so the graphs are bipartite. 

Moreover each Feynman graph has a boundary graph which is defined as follows: to each external leg of a Feynman graph is associated an external vertex so that the open graph is bipartite. These vertices are exactly the vertices of the boundary graph. An edge of colour $c$ in the boundary graph, corresponds to a path between two external legs in the Feynman graph, which alternates between dotted lines and lines of colour $c$. The boundary graphs are then $3$-coloured graphs, only composed of solid lines. A more detailed exposition of boundary graphs can be found in \cite{Perez-Sanchez:2016zbh} and \cite{Sanchez:2017gxt}.

The connected $2k$-point functions are then split into sectors indexed by a boundary graph $\mathcal{B}$, and taken to be
\begin{align}\label{def}
    \mathrm{G}_{\mathcal{B}}^{(2k)}\left(\mathbf{X}\right) =    
    \displaystyle \left. \frac{N^{-\alpha(\mathcal{B})}}{\mathrm{Z}_0} \prod_{i=1}^{k}\left(
    \frac{\delta}{\delta
      \bar{J}_{{\mathbf{p}^{i}}}}\frac{\delta}{\delta
      J_{\mathbf{x}^{i}}}\right)\mathrm{Z}[J,\bar{J}]\right|_{J=\bar{J}
      =0}\,,
\end{align}
where $Z_0=Z[0,0]$, ${\mathbf{X}} = (\mathbf{x}^1, \hdots, \mathbf{x}^k) \in \{ \frac{1}{N},\frac{2}{N}, \hdots, 1 \}^{3k}$ so that for all $c \in \{1,2,3\}$ and $(i,j) \in \{ 1, \hdots, k \}^2$, $x_{c}^{i} \neq x_{c}^{j}$. 

The 
$\mathbf{p}^i \in \{ \frac{1}{N},\frac{2}{N}, \hdots, 1 \}^3$ are momentum $3$-tuples depending on the coordinates $\mathbf{X}$ in a way constrained by the boundary graph $\B$. Hence the $2k$-point functions do not depend on the $\mathbf{p}^i$ but only on $\mathbf{X}$. For instance, for the boundary graph $V_1$ (see Fig. \ref{fig:boundary}), $\mathbb J(V_1)(\mathbf{x}^1,\mathbf{x}^2)=J_{\mathbf{x}^1} J_{\mathbf{x}^2}\bar{J}_{\mathbf{p}^1}
\bar{J}_{\mathbf{p}^2} 
=
J_{\mathbf{x}^1} J_{\mathbf{x}^2}\bar{J}_{x^1_1x^2_2x^2_3}\bar{J}_{x^2_1x^1_2x^1_3}$, where $\mathbf{p}^1=(x^1_1,x^2_2,x^2_3)$ and $\mathbf{p}^2=(x^2_1,x^1_2,x^1_3)$. %In the following and for lower point functions, we will prefer the simplified notation $\mathbf{x},\mathbf{y},\mathbf{z}$ instead of $\mathbf{x}^1,\mathbf{x}^2,\mathbf{x}^3$. 
Let us note that white and black vertices in a boundary graph $\mathcal{B}$, correspond in $\mathbb J(\mathcal{B})$ to the sources $J$ and $\bar{J}$ respectively. 

We also introduce the scalings $\alpha (\mathcal{B})$ for each boundary graph $\mathcal{B}$, note that they do not depend on the choice of colouring of the respective graph $\mathcal{B}$. For example, the three pillow graphs have the same scaling $\alpha(V_1)=\alpha (V_2)=\alpha (V_3)$.

The free energy is written as an expansion over boundary graphs (see again \cite{Perez-Sanchez:2016zbh} for more details):
\begin{equation} \label{eq:W_expansion}
    \mathrm{W}[J,\bar{J}] = \sum \limits_{k=1}^{\infty} \displaystyle\sum_{\substack{\mathcal{B}\in\partial_{\mathcal S_{\mathrm{int}}} \\
   V(\mathcal{B})=2k}} \sum \limits_{\mathbf{X}} \frac{N^{\alpha(\mathcal{B})}}{|\Aut(\mathcal{B})|}
    \mathrm{G}_{\mathcal{B}}^{(2k)}(\mathbf{X})\cdot\mathbb{J}(\mathcal{B})(\mathbf{X}),
\end{equation}
where $\partial_{\mathcal{S}_{\mathrm{int}}}$ is the set of boundary graphs associated to the interaction terms, $V(\mathcal{B})$ is the number of vertices of $\mathcal{B}$ and we note $\mathbb{J}(\mathcal{B})(\mathbf{X}) = J_{\mathbf{x}^1} \hdots J_{\mathbf{x}^k}\bar{J}_{\mathbf{p}^1} \hdots\bar{J}_{\mathbf{p}^k}$. Here $\Aut(\B)$ is the symmetry group of the graph $\mathcal{B}$,
which namely consists of all graph-automorphisms that preserve the bipartiteness 
in a strict sense --- black vertices are mapped to black vertices, white to white --- and respect the colour on edges (see
\cite[Def. 7 and examples]{Perez-Sanchez:2016zbh}).

%The free energy is written as an expansion over boundary graph (for details see \cite{Perez-Sanchez:2016zbh})
%\begin{equation} \label{eq:W_expansion}
%    \mathrm{W}[J,\bar{J}] = \sum \limits_{k=1}^{\infty} \displaystyle\sum_{\substack{\mathcal{B}\in\partial_{\mathcal S_{\mathrm{int}}} \\
%   V(\mathcal{B})=2k}} \sum \limits_{\mathbf{X}} \frac{N^{\alpha(\mathcal{B})}}{|\Aut(\mathcal{B})|}
%    \mathrm{G}_{\mathcal{B}}^{(2k)}(\mathbf{X})\cdot\mathbb{J}(\mathcal{B})(\mathbf{X}),
%\end{equation}
%where $\partial_{\mathcal{S}_{\mathrm{int}}}$ is the set of boundary graphs associated to the interaction terms, $V(\mathcal{B})$ is the number of vertices of $\mathcal{B}$, ${\mathbf{X}} = (\mathbf{x}^1, \hdots, \mathbf{x}^k) \in \{ \frac{1}{N},\frac{2}{N}, \hdots, 1 \}^{3k}$, $\Aut(\B)$ is the symmetry group of the graph $\mathcal{B}$,
%$\mathbb{J}(\mathcal{B})(\mathbf{X}) = J_{\mathbf{x}^1} \hdots J_{\mathbf{x}^k}\bar{J}_{\mathbf{p}^1} \hdots\bar{J}_{\mathbf{p}^k}$. Here 
%$\mathbf{p}^i=\mathbf{p}^i(\mathbf{X})\in \{ \frac{1}{N},\frac{2}{N}, \hdots, 1 \} ^3$ is a momentum triplet determined by the boundary graph $\B$. For instance, the boundary graph $V_1$ denoting the pillow graph for the colour 1 (see figure \ref{fig:boundary}), $\mathbb J(V_1)(\mathbf{x},\mathbf{y})=J_{\mathbf{x}} J_{\mathbf{y}}\bar{J}_{x_1y_2y_3}\bar{J}_{y_1x_2x_3}$.

%A $2k$-point function for a
%connected boundary graph $\mathcal{B}$ is 
%\begin{align}
%    \mathrm{G}_{\mathcal{B}}^{(2k)}\left(\mathbf{X}\right) =
%    \displaystyle \left. \frac{N^{-\alpha(\mathcal{B})}}{\mathrm{Z}_0} \prod_{i=1}^{k}\left(
%    \frac{\delta}{\delta
%      \bar{J}_{{\mathbf{p}^{i}}}}\frac{\delta}{\delta
%      J_{\mathbf{x}^{i}}}\right)\mathrm{Z}[J,\bar{J}]\right|_{J=\bar{J}
%      =0}\,,
%\end{align}

In \cite{Sanchez:2017gxt}, we used the Ward-Takahashi identity established in \cite{Perez-Sanchez:2016zbh} to determine the Schwinger-Dyson equation for $2k$-point function with a
connected boundary graph $\mathcal{B}$. Then in \cite{Pascalie:2018nrs}, requiring that the SDE for connected boundary graph have a well-defined large $N$ limit, we conjectured a general formula for the scalings
\begin{equation}
\alpha(\mathcal{B}) = 3-B-2g-2k,
\end{equation}
where $2k$ is the number of vertices of $\mathcal{B}$, $B$ is its number of connected components and $g$ its genus. The genus of a graph is the minimal integer $g$ such that the graph can be drawn without crossings on a surface of genus $g$.
%Let us mention here that the coefficient $\alpha (\mathcal{B})$ does not depend on the choice of colouring that can be made for the respective bubble. For the three pillow graphs one has: $\alpha(V_1)=\alpha (V_2)=\alpha (V_3).$ 
In particular, the SDE for the $2$-point function is
\begin{align}
   &\mathrm{G}^{(2)}(\mathbf{x}) = \frac{1}{|\mathbf{x}|^2} - \frac{2\lambda}{|\mathbf{x}|^2} \sum \limits_{a=1}^3 \Bigg(\frac{1}{N^2} \sum \limits_{\mathbf{q}_{\hat{a}}} \mathrm{G}^{(2)}(\mathbf{q}_{\hat{a}}x_a) \mathrm{G}^{(2)}(\mathbf{x}) +\frac{1}{N^4}\mathrm{G}^{(4)}_a(\mathbf{x},\mathbf{x})
   \nonumber\\
   &+ \frac{1}{N^5}\sum \limits_{\mathbf{q}_{\hat{a}}} \mathrm{G}^{(4)}_{\mathrm{m}|\mathrm{m}}(\mathbf{q}_{\hat{a}}x_a,\mathbf{x})
   + \frac{1}{N^2}\sum \limits_{q_a} \frac{\mathrm{G}^{(2)}(\mathbf{x}_{\hat{a}}q_a)-\mathrm{G}^{(2)}(\mathbf{x})}{x_a^2 - q_a^2} + \frac{1}{N^4}\sum\limits_{c\neq a} \sum \limits_{q_b}\mathrm{G}^{(4)}_{c}(\mathbf{x},\mathbf{x}_{\hat{b}}q_b) \Bigg),
\end{align}
where in the last term $b \neq c$ and $b \neq a$.
Following \cite{Pascalie:2018nrs}, for $N= \frac{\tilde{N}}{\Lambda}$ and using
\begin{equation}
    \lim_{\tilde{N} \to \infty}\frac{\Lambda}{\tilde{N}} \sum_{k=1}^{\tilde{N}} f\Big(\frac{k\Lambda}{\tilde{N}}\Big) = \int \limits_0^{\Lambda}\mathrm{d}x f(x),
\end{equation}
the SDE for the $2$-point function writes
\begin{align}\label{SDE2N}
    \mathrm{G}^{(2)}(\mathbf{x}) 
    = \left(1+|\mathbf{x}|^2 + 2\lambda\sum \limits_{c=1}^3\int \limits_0^{ \Lambda}\mathrm{d}\mathbf{q}_{\hat{c}} \mathrm{G}^{(2)}(\mathbf{q}_{\hat{c}}x_c) \right)^{-1},
\end{align}
where $\mathrm{d}\mathbf{q}_{\hat{c}} =  \prod_{d\neq c}\mathrm{d}q_d $. The aim of this paper is to solve the 2-point function of this type of models, in the limit $\Lambda = \infty$. We will mainly study the case with $c=1$ only, which is essential for solving the SDE, as we will see in the next section.

\section{Perturbative expansion}
\label{sec:3}

In this section we will to compute the first orders of the perturbative expansion of the 2-point function. We use a Taylor subtraction scheme to renormalize the UV divergences. For simplicity, let us plug in equation \eqref{SDE2N}, the following expansion of the 2-point function
\begin{equation}\label{expansion}
     \mathrm{G}^{(2)}(\mathbf{x}) = \sum \limits_{n \geq 0} \lambda^n \mathrm{G}^{(2)}_{n}(\mathbf{x}), 
\end{equation}
in order to obtain a recursive equation for $n \geq 1$, which writes:
\begin{equation}\label{rec}
    \mathrm{G}^{(2)}_n(\mathbf{x}) = -\frac{2}{|\mathbf{x}|^2+1}\sum\limits_{c=1}^3\int\mathrm{d}\mathbf{q}_{\hat{c}} \sum\limits_{k=0}^{n-1}\Big(\mathrm{G}^{(2)}_k(\mathbf{q}_{\hat{c}}x_c)-\frac{\delta_{k0}}{1+|\mathbf{q}_{\hat{c}}|^2}\Big)\mathrm{G}^{(2)}_{n-k-1}(\mathbf{x}),
\end{equation}
where $|\mathbf{q}_{\hat{c}}|^2 = \sum_{d\neq c}q_d^2$, the integration on $q_c$ is over $[0,\infty]$, and when $k=0$ we subtract the first Taylor term to regularise the divergent integration on the free propagator $G_0^{(2)}$.

\subsection{Model with the 3 quartic melonic interactions}\label{c=1}

\begin{figure}[ht]
    \centering
    \includegraphics[scale=0.25]{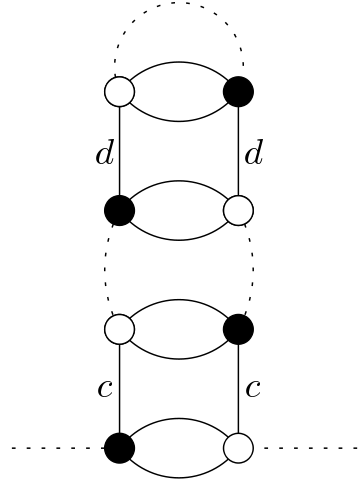}
    \caption{The only graphs at 2-loop order, giving contribution other than powers of logarithms for $d \neq c$.}
    \label{fig:pillowtower}
\end{figure}

Using the recursive equation~\eqref{rec}, we get
\allowdisplaybreaks
\begin{align}
    \mathrm{G}^{(2)}_{0}(\mathbf{x}) &= \frac{1}{1+|\mathbf{x}|^2}, \\
    \mathrm{G}^{(2)}_{1}(\mathbf{x}) &= -\frac{2}{(1+|\mathbf{x}|^2)^2}\sum_{c=1}^3\int \mathrm{d}\mathbf{q}_{\hat{c}}\Big(\frac{1}{1+|\mathbf{q}_{\hat{c}}x_c|^2}-\frac{1}{1+|\mathbf{q}_{\hat{c}}|^2}\Big) \nonumber \\
    &= \frac{\pi}{2(1+|\mathbf{x}|^2)^2}\sum_{c=1}^3\log{(x_c^2+1)}, \\
    \mathrm{G}^{(2)}_{2}(\mathbf{x}) &= -\frac{2}{1+|\mathbf{x}|^2}\sum_{c=1}^3\int \mathrm{d}\mathbf{q}_{\hat{c}}\Bigg\{\Big(\frac{1}{1+|\mathbf{q}_{\hat{c}}x_c|^2}-\frac{1}{1+|\mathbf{q}_{\hat{c}}|^2}\Big)\frac{\pi}{2(1+|\mathbf{x}|^2)^2}\sum_{d=1}^3\log{(x_d^2+1)} \nonumber \\
    &+ \frac{1}{1+|\mathbf{x}|^2}\frac{\pi}{2(1+|\mathbf{q}_{\hat{c}}x_c|^2)^2}\sum_{d=1}^3\log{((\mathbf{q}_{\hat{c}}x_c)_d+1)}\Bigg\}, \label{G2} \\
    &= \frac{1}{(1+|\mathbf{x}|^2)^2}\Bigg( \sum_{c=1}^3\sum_{d=1}^3\frac{\pi^2\log{(x_c^2+1)}\log{(x_d^2+1)}}{4(1+|\mathbf{x}|^2)}- \sum_{c=1}^3\frac{\pi\log{(x_c^2+1)}}{2(x_c^2+1)} \nonumber \\
    &-\pi^2\sum_{c=1}^3\frac{x_c \log \left(\frac{1}{4} \left(x_c^2+1\right)\right)+2 \tan ^{-1}(x_c)}{2 \left(x_c^3+x_c\right)} \Bigg).
\end{align}
We can remark that the last term in $\mathrm{G}^{(2)}_{2}(\mathbf{x})$ is the only term not containing powers of logarithms. It comes from the last term of~\eqref{G2} for $d \neq c$, which graphically corresponds to figure \ref{fig:pillowtower}.
This suggests that if we look at a model with only 1 pillow interaction, such graphs cannot exist, and the perturbative expansion should only be made of powers of logarithms.

\subsection{Model with 1 quartic melonic interaction}
\label{subsec:2}

Indeed, for only the pillow for the colour 1 as an interaction, we get:
\begin{equation}
    \mathrm{G}^{(2)}(\mathbf{x}) = \frac{1}{1+|\mathbf{x}|^2} +\frac{\pi\lambda}{2(1+|\mathbf{x}|^2)^2}\log{(x_1^2+1)} + \frac{(\pi\lambda)^2}{4(1+|\mathbf{x}|^2)^2}\Big( \frac{\log^2{(x_1^2+1)}}{(1+|\mathbf{x}|^2)}- \frac{\log{(x_1^2+1)}}{(x_1^2+1)} \Big) + O(\lambda^3).
\end{equation}
In this case, we can notice that only two types of integrals appear:
\begin{align}
    &\int \mathrm{d}\mathbf{q}_{\hat{1}}\Big(\frac{1}{1+|\mathbf{q}_{\hat{1}}x_1|^2}-\frac{1}{1+|\mathbf{q}_{\hat{1}}|^2}\Big) = -\frac{\pi}{4}\log{(x_1^2+1)}, \\
    &\int \mathrm{d}\mathbf{q}_{\hat{1}}\frac{1}{(1+|\mathbf{q}_{\hat{1}}x_1|^2)^n} = \frac{\pi(1+x_1^2)^{1-n}}{4(n-1)} \quad \text{for}\,\, n>1.
\end{align}
Hence, we can compute easily higher orders in the loop expansion\footnote{We computed the expansion up to order 9 in the coupling using Mathematica.}, which suggest the following form for all order $n$ in the coupling
\begin{equation}\label{ansatz}
    \mathrm{G}^{(2)}_{n}(\mathbf{x})= \bigg(\frac{\pi}{2}\bigg)^n\Bigg(\frac{\log^n(1+x_1^2)}{(1+|\mathbf{x}|^2)^{n+1}}+\frac{(-1)^n}{(1+x_1^2)^n}\sum\limits_{k=1}^{n-1}(-1)^k\log^k(1+x_1^2)\sum\limits_{m=1}^k a_{n,k,m}\frac{(1+x_1^2)^{m}}{(1+|\mathbf{x}|^2)^{m+1}}\Bigg),
\end{equation}
where  we conjecture that the numbers $a_{n,k,m}$ are
\begin{align}
    a_{n,k,m} &= \binom{n-1}{m-1} \frac{m!}{k!}|s_{n-m,n-k}| \nonumber \\
              &= (-1)^{k-m} (n-1)! m \frac{s_{n-m,n-k}}{(n-m)!k!}, 
\end{align}
where $s_{n,k}$ are the Stirling numbers of the 1st kind.
Using the change of variable $j=n-m$, we have
\begin{equation}
    b_{n,k,j} = (-1)^{k+j-n} (n-1)! (n-j) \frac{s_{j,n-k}}{j!k!}.
\end{equation}
Noting that $s_{j,n-k} = 0$ if $j<n-k$ and if $k=0$ or $k=n$, we can write the sum on $j$ from $1$ to $n-1$ and the sum on $k$ from $0$ to $n$. This leads to the following expression:
\begin{align}
    &\mathrm{G}^{(2)}_{n}(\mathbf{x})= \bigg(\frac{\pi}{2}\bigg)^n\Bigg(\frac{\log^n(1+x_1^2)}{(1+|\mathbf{x}|^2)^{n+1}}+\frac{(-1)^n}{(1+x_1^2)^n}\sum\limits_{k=1}^{n-1}(-1)^k\log^k(1+x_1^2)\sum\limits_{m=1}^k a_{n,k,m}\frac{(1+x_1^2)^{m}}{(1+|\mathbf{x}|^2)^{m+1}}\Bigg) \nonumber \\
    &= \bigg(\frac{\pi}{2}\bigg)^n\Bigg(\frac{\log^n(1+x_1^2)}{(1+|\mathbf{x}|^2)^{n+1}}+(n-1)!\sum\limits_{k=0}^{n}\sum\limits_{j=1}^{n-1} \frac{s_{j,n-k}}{j!k!}\frac{(-1)^j(n-j)}{(1+|\mathbf{x}|^2)^{n+1-j}(1+x_1^2)^j}\log^k(1+x_1^2)\Bigg). \label{Gn2}
\end{align}
The structure of the perturbative expansion is similar to the one studied in \cite{Panzer:2018tvy}. In the next section, we will sum the expansion following the same method.

\section{Resummation}
\label{sec:4}
In this section we perform the resummation of the perturbative expansion to obtain an explicit expression for the $2$-point function.
Let us use the formulas
\begin{align}
    &(-1)^{j}s_{j,n-k} = \frac{1}{(n-k)!}\frac{\mathrm{d}^{n-k}}{\mathrm{d}u^{n-k}}\frac{\Gamma(j-u)}{\Gamma(-u)}\Bigg|_{u=0}, \\
    &\log^k(1+x_1^2) =  \frac{\mathrm{d}^{k}}{\mathrm{d}u^{k}}(1+x_1^2)^u\Bigg|_{u=0},
\end{align}
to rewrite the second term of the RHS of \eqref{Gn2} as 
\begin{align}
&\bigg(\frac{\pi}{2}\bigg)^n\sum\limits_{j=1}^{n-1} \frac{n-j}{j!n}\frac{1}{(1+|\mathbf{x}|^2)^{n+1-j}(1+x_1^2)^j} \sum\limits_{k=0}^{n} \binom{n}{k}\Bigg(\frac{\mathrm{d}^{n-k}}{\mathrm{d}u^{n-k}}\frac{\Gamma(j-u)}{\Gamma(-u)}\Bigg) \Bigg(\frac{\mathrm{d}^{k}}{\mathrm{d}u^{k}}(1+x_1^2)^u\Bigg)\Bigg|_{u=0} \nonumber \\
&=\bigg(\frac{\pi}{2}\bigg)^n\sum\limits_{j=1}^{n-1} \frac{n-j}{j!n}\frac{1}{(1+|\mathbf{x}|^2)^{n+1-j}(1+x_1^2)^j} \frac{\mathrm{d}^{n}}{\mathrm{d}u^{n}}\Bigg(\frac{\Gamma(j-u)}{\Gamma(-u)}(1+x_1^2)^u\Bigg)\Bigg|_{u=0}.
\end{align}
Then using
\begin{align}
    \frac{\mathrm{d}^{n}}{\mathrm{d}u^{n}}\Bigg(\frac{\Gamma(j-u)}{\Gamma(-u)}(1+x_1^2)^u\Bigg) &= \frac{\mathrm{d}^{n}}{\mathrm{d}u^{n}}\Bigg((-1)^j(1+x_1^2)^j\frac{\mathrm{d}^{j}}{\mathrm{d}(x_1^2)^{j}}(1+x_1^2)^u\Bigg)\Bigg|_{u=0} \nonumber \\
    &= (-1)^j(1+x_1^2)^j\frac{\mathrm{d}^{j}}{\mathrm{d}(x_1^2)^{j}}\log^n(1+x_1^2), 
\end{align}
and realising that the first term of the rhs of \eqref{Gn2} corresponds to $j=0$, we have
\begin{align}
\mathrm{G}^{(2)}_{n}(\mathbf{x}) &= \bigg(\frac{\pi}{2}\bigg)^n\Bigg(\frac{\log^n(1+x_1^2)}{(1+|\mathbf{x}|^2)^{n+1}}+\sum\limits_{j=1}^{n-1} \frac{n-j}{j!n}\frac{(-1)^j}{(1+|\mathbf{x}|^2)^{n+1-j}} \frac{\mathrm{d}^{j}}{\mathrm{d}(x_1^2)^{j}}\log^n(1+x_1^2)\Bigg) \nonumber \\
&=\bigg(\frac{\pi}{2}\bigg)^n\sum\limits_{j=0}^{n-1} \frac{n-j}{j!n}\frac{(-1)^j}{(1+|\mathbf{x}|^2)^{n+1-j}} \frac{\mathrm{d}^{j}}{\mathrm{d}(x_1^2)^{j}}\log^n(1+x_1^2)\Bigg).
\end{align}
We then write
\begin{equation}
    \frac{1}{(1+|\mathbf{x}|^2)^{n+1-j}} = \frac{(-1)^{n-j}}{(n-j)!} \frac{\mathrm{d}^{n-j}}{\mathrm{d}(x_1^2)^{n-j}} \frac{1}{(1+|\mathbf{x}|^2)},
\end{equation}
to get
\begin{align}
\mathrm{G}^{(2)}(\mathbf{x})&=\frac{1}{1+|\mathbf{x}|^2} + \sum\limits_{n=1}^{\infty}\bigg(\frac{\pi}{2}\bigg)^n\frac{(-1)^n\lambda^n}{n!}\sum\limits_{j=0}^{n-1} \binom{n-1}{j} \frac{\mathrm{d}^{n-j}}{\mathrm{d}(x_1^2)^{n-j}} \frac{1}{(1+|\mathbf{x}|^2)} \frac{\mathrm{d}^{j}}{\mathrm{d}(x_1^2)^{j}}\log^n(1+x_1^2)\nonumber \\
&=\frac{1}{1+|\mathbf{x}|^2} - \sum\limits_{n=1}^{\infty}\bigg(\frac{\pi}{2}\bigg)^n\frac{\lambda^n}{n!}\frac{\mathrm{d}^{n-1}}{\mathrm{d}(x_1^2)^{n-1}} \frac{(-\log(1+x_1^2))^n}{(1+|\mathbf{x}|^2)^2}.
\end{align}
To sum this series, we use the Lagrange-B\"{u}rmann inversion formula \cite{Lagrange}, \cite{Burmann}. This formula states that for $\phi(\omega)$ analytic at $\omega = 0$, such that $\phi(0)\neq 0$ and $f(\omega) = \frac{\omega}{\phi(\omega)}$, the inverse function $g(z)$ of $f(\omega)$, such that $z=f(g(z))$, is analytic at $z=0$ and given by
\begin{equation}
    g(z)= \sum\limits_{n=1}^{\infty}\frac{z^n}{n!}\frac{\mathrm{d}^{n-1}}{\mathrm{d}\omega^{n-1}} \phi(\omega)^n\Bigg|_{\omega=0}. \label{LB1}
\end{equation}
Moreover, for any analytic function $H(z)$ such that $H(0)=0$, 
\begin{equation}
    H(g(z))= \sum\limits_{n=1}^{\infty}\frac{z^n}{n!}\frac{\mathrm{d}^{n-1}}{\mathrm{d}\omega^{n-1}}\bigg(H'(\omega) \phi(\omega)^n\bigg)\Bigg|_{\omega=0}. \label{LB2}
\end{equation}
Hence, for $z=\frac{\pi}{2}\lambda$, $\phi(\omega)=-\log(1+\omega+x_1^2)$ and $H(\omega)=\frac{1}{1+\omega+|\mathbf{x}|^2}-\frac{1}{1+|\mathbf{x}|^2}$, equation \eqref{LB1} gives
\begin{equation}
    g(x_1,z)= \sum\limits_{n=1}^{\infty}\frac{z^n}{n!}\frac{\mathrm{d}^{n-1}}{\mathrm{d}(x_1^2)^{n-1}}(-\log(1+x_1^2))^n,
\end{equation}
such that 
\begin{equation}\label{eqg}
    z= -\frac{ g(x_1,z)}{\log(1+g(x_1,z)+x_1^2)},
\end{equation}
which is solved by
\begin{equation}
    g(x_1,z) = z W\bigg(\frac{1}{z}e^{\frac{1+x_1^2}{z}}\bigg)-1-x_1^2,
\end{equation}
where $W(z)$ is the Lambert function defined by $z = W(ze^z)$.
Then, using equation \eqref{LB2}, we can write
\begin{align}\label{solution}
   \mathrm{G}^{(2)}(\mathbf{x})=\frac{1}{1+|\mathbf{x}|^2} -  \sum\limits_{n=1}^{\infty}\frac{z^n}{n!}\frac{\mathrm{d}^{n-1}}{\mathrm{d}(x_1^2)^{n-1}} \frac{(-\log(1+x_1^2))^n}{(1+|\mathbf{x}|^2)^2} = \frac{1}{1+|\mathbf{x}|^2+g(x_1,z)}.
\end{align}
This result can be integrated:
\begin{equation}
    \int \mathrm{d}\mathbf{q}_{\hat{1}} \big(G(\mathbf{q}_{\hat{c}}x_1)-\frac{1}{1+|\mathbf{q}_{\hat{1}}|^2}\big)
= -\frac{\pi}{4} \log\big(1+x_1^2 +g(x_1,z)\big).
\end{equation}
Using \eqref{SDE2N} for $c=1$, we recover \eqref{eqg}. 

We have thus proved that \eqref{solution} is a solution of the Schwinger-Dyson equation 
\begin{equation}
     \mathrm{G}^{(2)}(\mathbf{x})=\Big(1+|\mathbf{x}|^2 + 2\lambda\int \mathrm{d}\mathbf{q}_{\hat{1}} \big(G(\mathbf{q}_{\hat{c}}x_1)-\frac{1}{1+|\mathbf{q}_{\hat{1}}|^2}\big)\Big)^{-1}. 
\end{equation}
In the limit $\lambda \rightarrow 0$, using $W(x) = \log x - \log \log x + o(1)$ we get 
\begin{equation}
    \lim_{\lambda \to 0} \frac{\pi\lambda}{2} W\Big(\frac{2}{\pi\lambda}e^{\frac{2(1+x_1^2)}{\pi\lambda}}\Big) = 1 +x_1^2,
\end{equation}
so that
\begin{equation}
     \lim_{\lambda \to 0} \mathrm{G}^{(2)}(\mathbf{x})=\frac{1}{1+|\mathbf{x}|^2},
\end{equation}
and we recover the free propagator, as expected.

We established our solution for $\lambda > 0$ with $x_1,x_2,x_3 \geq 0$ but it can be analytically extended. The Lambert function has many branches behaving differently on the complex plane \cite{Corless1996}, the branch assignment of our solution depends on $\lambda$. We give a short comment on the holomorphic extension of our solution in $z=\frac{\pi}{2}\lambda$ for a fixed $\mathbf{x}$, which is heavily based on \cite{Panzer:2018tvy} where all the details are discussed.

One first needs to study the map $z\rightarrow zW\Big(\frac{1}{z}e^{\frac{1+x_1^2}{z}}\Big) $ to get proposition 15 of \cite{Panzer:2018tvy}. Taking into account a rescaling of $\frac{2}{\pi}$ to translate their results in term of our $\lambda$, this map is holomorphic on $\mathbb{C}\backslash\{-\frac{2}{\pi}(1+x_1^2)\frac{\sin{\alpha}}{\alpha}| \frac{\sin{\alpha}}{\alpha}e^{\alpha \cot{\alpha}}\geq \frac{\pi e}{2(1+x_1^2)}, -\pi < \alpha < \pi\}$. Varying $x_1$, the common holomorphic domain $\Omega$ is at the right of the curve $\mathcal{C}=\{-e^{1-\alpha \cot{\alpha}+i\alpha}| -\pi < \alpha < \pi\}$ and is not affected by the rescaling. In particular, it contains the disk $|\lambda|<1$ and the map has a convergent radius $\geq 1$ in $\lambda$ for all $x_1 \geq 0$. In our case, we can have poles if $ zW\Big(\frac{1}{z}e^{\frac{1+x_1^2}{z}}\Big) = -x_2^2-x_3^2 \mp i\epsilon = -y \mp i\epsilon$ with $y>0$. This equation can be solved with $\epsilon \rightarrow 0$ by $\lambda^{\pm}_{x_1}(y)$, a critical line in the $z$-plane, parametrised by $y$ ($\lambda^{\pm}_{a}(\varphi)$ in the notation of \cite{Panzer:2018tvy}, Lemma 18) and with a specific branch of the Lambert function. For this branch we would get a pole, however the actual branch assignment in our solution for $\lambda = \lambda^{\pm}_{x_1}(y)$ is a different branch and the critical line does not cause any singularity. The two-point function is then holomorphic in the domain $\Omega$ of the complex plane depicted in figure 1 of \cite{Panzer:2018tvy}.

%In the strong coupling limit, using $\frac{W(x)}{x} = 1 - \frac{1}{x} + O(x) $, we get
%\begin{equation}
%    \frac{\pi\lambda}{2} W\Big(\frac{2}{\pi\lambda}e^{\frac{2(1+x_1^2)}{\pi\lambda}}\Big) = 1 + \frac{2}{\pi\lambda} + O\Big(\frac{1}{\lambda^2}\Big).
%\end{equation}
%Then
%\begin{equation}
%     \mathrm{G}^{(2)}(\mathbf{x})=\frac{1}{1+|\mathbf{x}|^2} - \log\Big( 1 + \frac{\frac{2}{\pi\lambda}-x_1^2}{(1+|\mathbf{x}|^2)^2}\Big),
%\end{equation}
%and equation \eqref{SDE2N} with only $c=1$ becomes
%\begin{equation}
%    \int\mathrm{d}q_2\mathrm{d}q_3 \mathrm{G}^{(2)}(x_1,q_2,q_3) = -\frac{1+|\mathbf{x}|^2}{2\lambda}.
%\end{equation}

\section{Higher-point functions}
\label{sec:5}

\begin{figure} 
\centering
\includegraphics[scale=1]{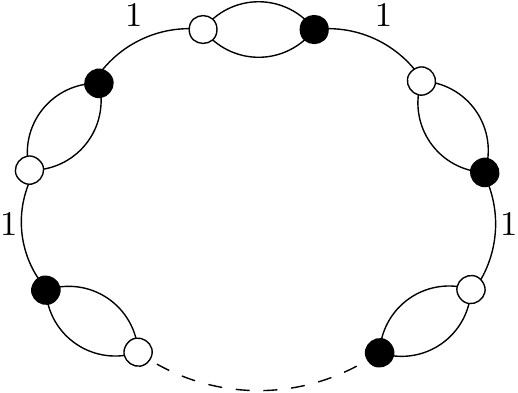}
\caption{General form of the connected components of boundary graphs in the model with 1 quartic melonic interaction. \label{fig:hboundary}}
\end{figure}

The boundary graphs of the model with 1 quartic melonic interaction have connected components of the form of figure \ref{fig:hboundary}. The $2k$-point function SDE with connected boundary graph was derived in the section $6$ of \cite{Sanchez:2017gxt}, taking the large $N$ limit established in \cite{Pascalie:2018nrs} and explained in section \ref{sec:2}, we get
\begin{align}
    G^{(2k)}(\mathbf{X}) &= 2\lambda G^{(2)}(x_1^1,x_2^2,x_3^2)\nonumber\\
    &\times\sum \limits_{\rho=2}^kG^{(2k-2\rho+2)}(\mathbf{x}^{\rho},\hdots,\mathbf{x}^{k})\frac{G^{(2\rho-2)}(\mathbf{x}^{1},\hdots,\mathbf{x}^{\rho-1})-G^{(2\rho-2)}(x^{\rho}_1,x_2^1,x_3^1,\hdots,\mathbf{x}^{\rho-1})}{(x_1^1)^2-(x_1^{\rho})^2}.
\end{align}
From the solution \eqref{solution} of the 2-point function SDE, we can recursively obtain any higher-point function with a connected boundary graph.

The case of disconnected boundary graph is more involved \cite{Perez-Sanchez:2018qkd} and no general expression of the SDE in the large $N$ limit have yet been obtained. The simplest equation is the SDE for the $4$-point function with disconnected boundary graph, which for only 1 quartic interaction and in the large $N$ limit reduces to 
\begin{equation}\label{4ptSDE}
    \mathrm{G}^{(4)}_{\mathrm{m}|\mathrm{m}}(\mathbf{x},\mathbf{y}) =- 2\lambda(\mathrm{G}^{(2)}(\mathbf{x}))^2\int\mathrm{d}q_2\mathrm{d}q_3\mathrm{G}^{(4)}_{\mathrm{m}|\mathrm{m}}(x_1,q_2,q_3,\mathbf{y}).
\end{equation}
Analysing the perturbative expansion of $\mathrm{G}^{(4)}_{\mathrm{m}|\mathrm{m}}$, we see that there is no contribution at order $\lambda^ 0$, since the Feynman graph which can contribute (made with two free propagator) is disconnected.
Moreover in the appendix of \cite{Pascalie:2018nrs}, we determined that the first contribution to the perturbative expansion is at order $\lambda^ 2$ and corresponds to graphs built with 2 different pillow interactions, of the form of the Feynman graph of figure \ref{fig:boundary} b). In the present case of the model with only 1 quartic melonic interaction, no such graph exists. Then, by plugging an expansion of the form of \eqref{expansion} for the $2$- and $4$-point functions in \eqref{4ptSDE}, we can recursively establish that all order of the perturbative expansion of $\mathrm{G}^{(4)}_{\mathrm{m}|\mathrm{m}}$ are null. Hence, at leading order in the large $N$ limit, $\mathrm{G}^{(4)}_{\mathrm{m}|\mathrm{m}}$ is completely suppressed.

\section{Concluding remarks}
\label{sec:6}

In this paper we have solved the $2$-point function of a tensor field theory with 1 quartic melonic interaction, with building block the Lambert-W function, using a perturbative expansion and a Lagrange-B\"{u}rmann resummation. From this result, all higher-point functions with connected boundary graph can be obtained recursively. Moreover, we have shown by a perturbative argument that the $4$-point function with a disconnected boundary graph is null at leading order in the large $N$ limit. 

\medskip

A first perspective for this work is the study of higher-point functions with disconnected boundary graph. As mention in section \ref{sec:2}, the form of the scalings $\alpha(\mathcal{B})$ is a conjecture made in \cite{Pascalie:2018nrs} and based on the study of SDE with connected boundary graph. Now one has to prove this conjecture using the SDE for disconnected boundary graph that have been determined in \cite{Perez-Sanchez:2018qkd}. As in the connected boundary case, the large $N$ limit of these SDE are expected to involve only lower point function and in particular the $2$-point function found in this paper. The fact that the $4$-point function with a disconnected boundary graph is null may indicate that at least some of the higher-point functions will also be suppressed at leading order in $N$.

\medskip

Another perspective which appears interesting to us is the study the model with 3 quartic melonic interactions. The perturbative expansion is more involved but other techniques such as the blobbed topological expansion for such tensor model \cite{Bonzom:2016kqf} may prove useful.  

\section*{Acknowledgement}

The author would like to thank Raimar Wulkenhaar for his guidance throughout this project, Adrian Tanasa for his advice and comments on the manuscript, and Alexander Hock for helpful discussions. %Funded by the Deutsche Forschungsgemeinschaft (DFG, German Research
%Foundation) under Germany's Excellence Strategy EXC 2044--390685587,
%Mathematics Münster: Dynamics--Geometry--Structure. This research has also been partially supported by the CNRS Infiniti ModTens grant.

\appendix %Passage au mode appendice
\renewcommand{\thesection}{\Alph{section}} %On change la numérotation pour avoir des lettres

%\vfill\eject  
\section{Recurrence relations}
\label{appendix:1}

In this section, we will use the recursive equation \eqref{rec} to determine recurrence relations on the numbers $a_{n,k,m}$. We first perform the integration
\begin{align}
&\int\mathrm{d}\mathbf{q}_{\hat{1}} \mathrm{G}^{(2)}_p(\mathbf{q}_{\hat{1}}x_1) = -\frac{\pi}{4}\log(1+x_1^2) \quad \text{if} \,\, p=0,\\
&= \bigg(\frac{\pi}{2}\bigg)^{p+1}\Bigg(\frac{\log^p(1+x_1^2)}{2p(1+x_1^2)^p}+\frac{(-1)^p}{2(1+x_1^2)^p}\sum\limits_{r=1}^{p-1}(-1)^r\log^r(1+x_1^2)\sum\limits_{m=1}^{r}\frac{a_{p,r,m}}{m} \Bigg) \quad \text{if} \,\, p>0,
\end{align}
where for $p=1$ the sum on $r$ does not appear. Plugging back the ansatz \eqref{ansatz} in the recurrence relation~\eqref{rec} with $c=1$ gives
\begin{align}\label{rec1}
   &\mathrm{G}^{(2)}_{n}(\mathbf{x})= -\frac{2}{|\mathbf{x}|^2+1}\Bigg\{-\frac{\pi}{4}\log(1+x_1^2)\mathrm{G}^{(2)}_{n-1}(\mathbf{x}) + \sum\limits_{p=1}^{n-1}\bigg(\frac{\pi}{2}\bigg)^{p+1}\frac{\log^p(1+x_1^2)}{2p(1+x_1^2)^p} \mathrm{G}^{(2)}_{n-p-1}(\mathbf{x}) \nonumber\\ 
   &+ \sum\limits_{p=2}^{n-1}\bigg(\frac{\pi}{2}\bigg)^{p+1}\bigg(\frac{(-1)^p}{2(1+x_1^2)^p}\sum\limits_{r=1}^{p-1}(-1)^r\log^r(1+x_1^2)\sum\limits_{m=1}^{r}\frac{a_{p,r,m}}{m} \Bigg)\mathrm{G}^{(2)}_{n-p-1}(\mathbf{x})\Bigg\}.
\end{align}
The first term of \eqref{rec1} gives
\begin{align}
    &\frac{\pi\log(1+x_1^2)}{2(|\mathbf{x}|^2+1)}\bigg(\frac{\pi}{2}\bigg)^{n-1}\Bigg(\frac{\log^{n-1}(1+x_1^2)}{(1+|\mathbf{x}|^2)^{n}}+\frac{(-1)^{n-1}}{(1+x_1^2)^{n-1}}\sum\limits_{k=1}^{n-2}(-1)^k\log^k(1+x_1^2)\sum\limits_{m=1}^k a_{n-1,k,m}\frac{(1+x_1^2)^{m}}{(1+|\mathbf{x}|^2)^{m+1}}\Bigg) \nonumber\\
    &=\bigg(\frac{\pi}{2}\bigg)^{n}\Bigg(\frac{\log^{n}(1+x_1^2)}{(1+|\mathbf{x}|^2)^{n+1}}+\frac{(-1)^{n-1}}{(1+x_1^2)^{n-1}}\sum\limits_{k=1}^{n-2}(-1)^k\log^{k+1}(1+x_1^2)\sum\limits_{m=1}^k a_{n-1,k,m}\frac{(1+x_1^2)^{m}}{(1+|\mathbf{x}|^2)^{m+2}}\Bigg) \nonumber \\
    &=\bigg(\frac{\pi}{2}\bigg)^{n}\Bigg(\frac{\log^{n}(1+x_1^2)}{(1+|\mathbf{x}|^2)^{n+1}}+\frac{(-1)^{n}}{(1+x_1^2)^{n}}\sum\limits_{k=2}^{n-1}(-1)^k\log^{k}(1+x_1^2)\sum\limits_{m=2}^k a_{n-1,k-1,m-1}\frac{(1+x_1^2)^{m}}{(1+|\mathbf{x}|^2)^{m+1}}\Bigg),
\end{align}
where we sent $k \rightarrow k+1$ and $m \rightarrow m+1$ to get to the last line.
The second term of~\eqref{rec1} gives
\begin{align}
    &-\frac{2}{|\mathbf{x}|^2+1}\Bigg(\sum\limits_{p=1}^{n-1}\bigg(\frac{\pi}{2}\bigg)^{p+1}\frac{\log^p(1+x_1^2)}{2p(1+x_1^2)^p} \bigg(\frac{\pi}{2}\bigg)^{n-p-1}\frac{\log^{n-p-1}(1+x_1^2)}{(1+|\mathbf{x}|^2)^{n-p}} \nonumber\\
    &+\sum\limits_{p=1}^{n-3}\bigg(\frac{\pi}{2}\bigg)^{p+1}\frac{\log^p(1+x_1^2)}{2p(1+x_1^2)^p}\bigg(\frac{\pi}{2}\bigg)^{n-p-1}\frac{(-1)^{n-p-1}}{(1+x_1^2)^{n-p-1}}\sum\limits_{k=1}^{n-p-2}(-1)^k\log^k(1+x_1^2)\sum\limits_{m=1}^k a_{n-p-1,k,m}\frac{(1+x_1^2)^{m}}{(1+|\mathbf{x}|^2)^{m+1}}\Bigg) \nonumber\\
    &=-\bigg(\frac{\pi}{2}\bigg)^{n}\frac{\log^{n-1}(1+x_1^2)}{(1+x_1^2)^{n}} \sum\limits_{p=1}^{n-1}\frac{1}{p}\frac{(1+x_1^2)^{n-p}}{(1+|\mathbf{x}|^2)^{n-p+1}}
    \nonumber\\
    &+\bigg(\frac{\pi}{2}\bigg)^{n}\frac{(-1)^n}{(1+x_1^2)^n}\sum\limits_{p=1}^{n-3}\sum\limits_{k=1}^{n-p-2}(-1)^{k-p}\log^{p+k}(1+x_1^2)\sum\limits_{m=1}^k \frac{a_{n-p-1,k,m}}{p}\frac{(1+x_1^2)^{m+1}}{(1+|\mathbf{x}|^2)^{m+2}}. 
\end{align}
Setting $r=p+k$ in the line of the previous equation, let us rewrite the double sum as
\begin{equation}
    \sum\limits_{k=1}^{n-3}\sum\limits_{r=k+1}^{n-2}\frac{(-1)^{r}}{r-k}\log^{r}(1+x_1^2)a_{n-r+k-1,k,m} = \sum\limits_{r=2}^{n-2}\sum\limits_{k=1}^{r-1}\frac{(-1)^{r}}{r-k}\log^{r}(1+x_1^2)a_{n-r+k-1,k,m}.
\end{equation}
Then we send $m \rightarrow m+1$ and rewrite double sum to get
\begin{align}
\sum\limits_{k=1}^{r-1}\sum\limits_{m=2}^{k+1} \frac{a_{n-r+k-1,k,m-1}}{r-k}\frac{(1+x_1^2)^{m}}{(1+|\mathbf{x}|^2)^{m+1}} = \sum\limits_{m=2}^{r}\sum\limits_{k=m-1}^{r-1} \frac{a_{n-r+k-1,k,m-1}}{r-k}\frac{(1+x_1^2)^{m}}{(1+|\mathbf{x}|^2)^{m+1}}. 
\end{align}
Hence, sending $p \rightarrow n-p$ and collecting the results we get
\begin{align}
    &-\bigg(\frac{\pi}{2}\bigg)^{n}\frac{\log^{n-1}(1+x_1^2)}{(1+x_1^2)^{n}} \sum\limits_{p=1}^{n-1}\frac{1}{n-p}\frac{(1+x_1^2)^{p}}{(1+|\mathbf{x}|^2)^{p+1}}
    \nonumber\\
    &+\bigg(\frac{\pi}{2}\bigg)^{n}\frac{(-1)^n}{(1+x_1^2)^n}\sum\limits_{r=2}^{n-2}(-1)^{r}\log^{r}(1+x_1^2)\sum\limits_{m=2}^{r}\sum\limits_{k=m-1}^{r-1} \frac{a_{n-1+k-r,k,m-1}}{r-k}\frac{(1+x_1^2)^{m}}{(1+|\mathbf{x}|^2)^{m+1}}. 
\end{align}
The third term of~\eqref{rec1} gives
\begin{align}\label{rec2}
    &-\frac{2}{|\mathbf{x}|^2+1}\Bigg\{\sum\limits_{p=2}^{n-1}\bigg(\frac{\pi}{2}\bigg)^{p+1}\Bigg(\frac{(-1)^p}{2(1+x_1^2)^p}\sum\limits_{r=1}^{p-1}(-1)^r\log^r(1+x_1^2)\sum\limits_{m=1}^{r}\frac{a_{p,r,m}}{m} \Bigg)\bigg(\frac{\pi}{2}\bigg)^{n-p-1}\frac{\log^{n-p-1}(1+x_1^2)}{(1+|\mathbf{x}|^2)^{n-p}} \nonumber\\
    &+\sum\limits_{p=2}^{n-3}\bigg(\frac{\pi}{2}\bigg)^{p+1}\Bigg(\frac{(-1)^p}{2(1+x_1^2)^p}\sum\limits_{r=1}^{p-1}(-1)^r\log^r(1+x_1^2)\sum\limits_{m=1}^{r}\frac{a_{p,r,m}}{m} \Bigg) \nonumber\\
    &\bigg(\frac{\pi}{2}\bigg)^{n-p-1}\frac{(-1)^{n-p-1}}{(1+x_1^2)^{n-p-1}}\sum\limits_{k=1}^{n-p-2}(-1)^k\log^k(1+x_1^2)\sum\limits_{l=1}^k a_{n-p-1,k,l}\frac{(1+x_1^2)^{l}}{(1+|\mathbf{x}|^2)^{l+1}}\Bigg)\Bigg\}. 
\end{align}
The first term of equation \eqref{rec2} gives
\begin{align}
  &-\bigg(\frac{\pi}{2}\bigg)^{n}  \sum\limits_{p=2}^{n-1}\sum\limits_{r=1}^{p-1}(-1)^{p+r}\frac{\log^{n-p+r-1}(1+x_1^2)}{(1+x_1^2)^n}\sum\limits_{m=1}^{r}\frac{a_{p,r,m}}{m}\frac{(1+x_1^2)^{n-p}}{(1+|\mathbf{x}|^2)^{n-p+1}} \nonumber \\
  &=-\bigg(\frac{\pi}{2}\bigg)^{n} \sum\limits_{r=1}^{n-2}\sum\limits_{k=1}^{n-1-r}(-1)^k\frac{\log^{n-k-1}(1+x_1^2)}{(1+x_1^2)^n}\sum\limits_{m=1}^{r}\frac{a_{r+k,r,m}}{m}\frac{(1+x_1^2)^{n-r-k}}{(1+|\mathbf{x}|^2)^{n-r-k+1}}, 
\end{align}
by setting $k=p-r$. Then by setting $l=n-1-k$ and rewriting the sums we get
\begin{align}
  &\bigg(\frac{\pi}{2}\bigg)^{n}\frac{(-1)^n}{(1+x_1^2)^n}\sum\limits_{r=1}^{n-2}\sum\limits_{l=r}^{n-2}(-1)^l\log^{l}(1+x_1^2)\sum\limits_{m=1}^{r}\frac{a_{n-1+r-l,r,m}}{m}\frac{(1+x_1^2)^{l-r+1}}{(1+|\mathbf{x}|^2)^{l-r+2}} \nonumber \\
  &=\bigg(\frac{\pi}{2}\bigg)^{n}\frac{(-1)^n}{(1+x_1^2)^n}\sum\limits_{l=1}^{n-2}(-1)^l\log^{l}(1+x_1^2)\sum\limits_{r=1}^{l}\sum\limits_{m=1}^{r}\frac{a_{n-1+r-l,r,m}}{m}\frac{(1+x_1^2)^{l-r+1}}{(1+|\mathbf{x}|^2)^{l-r+2}}. 
\end{align}
Then we set $k = l-r+1$ and obtain
\begin{equation}
    \bigg(\frac{\pi}{2}\bigg)^{n}\frac{(-1)^n}{(1+x_1^2)^n}\sum\limits_{l=1}^{n-2}(-1)^l\log^{l}(1+x_1^2)\sum\limits_{k=1}^{l}\sum\limits_{m=1}^{l-k+1}\frac{a_{n-k,l-k+1,m}}{m}\frac{(1+x_1^2)^{k}}{(1+|\mathbf{x}|^2)^{k+1}}. 
\end{equation}
The second term of~\eqref{rec2} gives, by rewriting the sums,
\begin{align}
    &\bigg(\frac{\pi}{2}\bigg)^{n}\frac{(-1)^n}{(1+x_1^2)^n}\sum\limits_{p=2}^{n-3}\sum\limits_{k=1}^{n-p-2}\sum\limits_{r=1}^{p-1}(-1)^{k+r}\log^{k+r}(1+x_1^2)\sum\limits_{l=1}^k\sum\limits_{m=1}^{r}\frac{a_{p,r,m}}{m} a_{n-p-1,k,l}\frac{(1+x_1^2)^{l+1}}{(1+|\mathbf{x}|^2)^{l+2}} \nonumber \\
    &=\bigg(\frac{\pi}{2}\bigg)^{n}\frac{(-1)^n}{(1+x_1^2)^n}\sum\limits_{r=1}^{n-4}\sum\limits_{k=1}^{n-3-r}(-1)^{k+r}\log^{r+k}(1+x_1^2)\sum\limits_{l=1}^{k}\sum\limits_{m=1}^{r}\sum\limits_{p=r+1}^{n-2-k}\frac{a_{p,r,m}}{m} a_{n-p-1,k,l}\frac{(1+x_1^2)^{l+1}}{(1+|\mathbf{x}|^2)^{l+2}}.
\end{align}
First by setting $q=k+r$ and by several rewriting of the sums we get
\begin{align}
    &\bigg(\frac{\pi}{2}\bigg)^{n}\frac{(-1)^n}{(1+x_1^2)^n}\sum\limits_{k=1}^{n-4}\sum\limits_{q=k+1}^{n-3}(-1)^q\log^{q}(1+x_1^2)\sum\limits_{l=1}^{k}\sum\limits_{m=1}^{q-k}\sum\limits_{p=q-k+1}^{n-2-k}\frac{a_{p,q-k,m}}{m} a_{n-p-1,k,l}\frac{(1+x_1^2)^{l+1}}{(1+|\mathbf{x}|^2)^{l+2}} \nonumber \\
    &=\bigg(\frac{\pi}{2}\bigg)^{n}\frac{(-1)^n}{(1+x_1^2)^n}\sum\limits_{q=2}^{n-3}(-1)^q\log^{q}(1+x_1^2)\sum\limits_{k=1}^{q-1}\sum\limits_{l=1}^{k}\sum\limits_{m=1}^{q-k}\sum\limits_{p=q-k+1}^{n-2-k}\frac{a_{p,q-k,m}}{m} a_{n-p-1,k,l}\frac{(1+x_1^2)^{l+1}}{(1+|\mathbf{x}|^2)^{l+2}} \nonumber \\
    &=\bigg(\frac{\pi}{2}\bigg)^{n}\frac{(-1)^n}{(1+x_1^2)^n}\sum\limits_{q=2}^{n-3}(-1)^q\log^{q}(1+x_1^2)\sum\limits_{l=1}^{q-1}\sum\limits_{k=l}^{q-1}\sum\limits_{m=1}^{q-k}\sum\limits_{p=q-k+1}^{n-2-k}\frac{a_{p,q-k,m}}{m} a_{n-p-1,k,l}\frac{(1+x_1^2)^{l+1}}{(1+|\mathbf{x}|^2)^{l+2}} \nonumber \\
    &=\bigg(\frac{\pi}{2}\bigg)^{n}\frac{(-1)^n}{(1+x_1^2)^n}\sum\limits_{q=2}^{n-3}(-1)^q\log^{q}(1+x_1^2)\sum\limits_{l=2}^{q}\sum\limits_{k=l-1}^{q-1}\sum\limits_{m=1}^{q-k}\sum\limits_{p=q-k+1}^{n-2-k}\frac{a_{p,q-k,m}}{m} a_{n-p-1,k,l-1}\frac{(1+x_1^2)^{l}}{(1+|\mathbf{x}|^2)^{l+1}},
\end{align}
where we send $l \rightarrow l+1$ in the last line. 

\allowdisplaybreaks
Now collecting all the results we obtain recurrence relations on $a_{n,k,m}$:
\begin{align}
    &a_{n,1,1} = a_{n-1,1,1}, \\
    &a_{n,n-1,1} = \frac{1}{n-1}, \\
    &a_{n,n-1,m} =  \frac{1}{n-m} + a_{n-1,n-2,m-1}, \quad \text{for} \,\, m \in \llbracket 2,n-1 \rrbracket, \\
    &a_{n,n-2,m} = a_{n-1,n-3,m-1} + \sum\limits_{r=m-1}^{n-3} \frac{a_{r+1,r,m-1}}{n-2-r} + \sum\limits_{l=1}^{n-1-m} \frac{a_{n-m,n-1-m,l}}{l}, \label{recrel2}\\ 
    &\text{for} \,\, m \in \llbracket 2,n-2 \rrbracket, \nonumber  \\
    &a_{n,k,1} = \sum\limits_{l=1}^{k} \frac{a_{n-1,k,l}}{l}, \quad \text{for} \,\, k \in \llbracket 1,n-3 \rrbracket, \label{recrel1}\\
    &a_{n,k,m} = a_{n-1,k-1,m-1} + \sum\limits_{r=m-1}^{k-1} \frac{a_{n-1+r-k,r,m-1}}{k-r} + \sum\limits_{l=1}^{k-m+1} \frac{a_{n-m,k-m+1,l}}{l} \nonumber \\ 
    &+ \sum\limits_{r=m-1}^{k-1}\sum\limits_{l=1}^{k-r}\sum\limits_{p=k-r+1}^{n-2-r}\frac{a_{p,k-r,l}a_{n-p-1,r,m-1}}{l}, 
    \quad \text{for} \,\, k \in \llbracket 2,n-3 \rrbracket \,\, \text{and} \,\, m \in \llbracket 2,k \rrbracket. \label{recrel3}
\end{align}
Rewriting these equations gives explicit relations on Stirling numbers of the first kind, harmonic numbers and binomial coefficients. Indeed, from equation \eqref{recrel1} we recover
\begin{equation}
   \frac{1}{(n-1)!} \stirlingi{n-1}{n-k} = \sum_{l=1}^k \frac{1}{(n-l)!} \stirlingi{n-1-l}{n-1-k}, \quad \text{for} \,\, k \in \llbracket 1,n-3 \rrbracket,
\end{equation}
which correspond to the equation (6.21) in \cite{Graham:1994:CMF:562056}.
%\begin{equation}
%  H_{n-m-1} = \sum\limits_{r=m-1}^{n-3} \frac{n-m}{(n-2-r)(r-m+2)} + \sum\limits_{l=1}^{n-1-m} \frac{n-m}{(m-1)(n-m-l)},
%\end{equation}
%for $m \in \llbracket 2,n-2 \rrbracket$. Then, we send $m \rightarrow n-k-1$ to have
%\begin{equation}
%    H_k = \sum\limits_{r=n-2-k}^{n-3} \frac{k+1}{(n-2-r)(r+k+3)} + \sum\limits_{l=1}^{k} \frac{k+1}{(n-k-2)(k+1-l)},
%\end{equation}
%for $k \in \llbracket 1,n-3 \rrbracket$. 
Setting $l=n-2-r $, $k = n-m-1 $ and sending $n-3 \rightarrow n$, equation \eqref{recrel2} gives
\begin{equation}
     H_k = \frac{k+1}{2n+3-k}\sum\limits_{l=1}^{k} \frac{n+1-k+l}{l(k+1-l)}, \quad \text{for} \,\, k \in \llbracket 1,n \rrbracket.
\end{equation}
Sending $r \rightarrow k-l$ and in the last term $l \rightarrow r$ of equation \eqref{recrel3}, we get
\begin{align}
    &((n-1)m-k(m-1))\frac{(n-2)!}{k!(n-m)!}\stirlingi{n-m}{n-k} \nonumber \\
    &= \sum \limits_{l=1}^{k-m+1}  \frac{1}{(n-m-l)!}\stirlingi{n-m-l}{n-k-1}\Bigg(\frac{(n-1-m)!}{(k-m+1)!}+\frac{m-1}{l}\frac{(n-l-2)!}{(k-l)!} \Bigg) \nonumber \\
    &+ \sum \limits_{l=1}^{k-m+1} \frac{m-1}{l!(k-l)!} \sum\limits_{p=l+1}^{n-2-k+l}\frac{(p-1)!(n-2-p)!}{(n-m-p)!}\stirlingi{n-m-p}{n-k-1-p+l}\sum_{r=1}^l\frac{1}{(p-r)!}\stirlingi{p-r}{p-l},
\end{align}
for $k \in \llbracket 2,n-3 \rrbracket$ and $m \in \llbracket 2,k \rrbracket$.

\printbibliography

\end{document}